\documentclass[aps,pra,groupaddress,showpacs,twocolumn]{revtex4-1}
\usepackage{graphicx,bm,amsmath,amsfonts,bbold, amssymb,color,microtype}

\newcommand{\be}{\begin{equation}}
\newcommand{\ee}{\end{equation}}
\newcommand{\bea}{\begin{eqnarray}}
\newcommand{\eea}{\end{eqnarray}}
\newcommand{\im}[1]{\mbox{Im}[#1]}
\newcommand{\re}[1]{\mbox{Re}[#1]}

\begin{document}

\title{Optical Reciprocity Induced Symmetry of the Scattering Eigenstates in Non-$\cal PT$-Symmetric Heterostructures}

\author{Li Ge}
\email{li.ge@csi.cuny.edu}
\affiliation{\textls[-18]{Department of Engineering Science and Physics, College of Staten Island, CUNY, Staten Island, NY 10314, USA}}
\affiliation{The Graduate Center, CUNY, New York, NY 10016, USA}
\author{Liang Feng}
\affiliation{Department of Electrical Engineering, The State University of New York at Buffalo, Buffalo, NY 14260, USA}

\begin{abstract}
The scattering matrix $S$ obeys the unitary relation $S^\dagger S=1$ in a Hermitian system and the symmetry property ${\cal PT}S{\cal PT}=S^{-1}$ in a Parity-Time (${\cal PT}$) symmetric system. Here we report a different symmetry relation of the $S$ matrix in a one-dimensional heterostructure, which is given by the amplitude ratio of the incident waves in the scattering eigenstates. It originates from the optical reciprocity and holds independent of the Hermiticity or $\cal PT$ symmetry of the system. Using this symmetry relation, we probe a quasi-transition that is reminiscent of the spontaneous symmetry breaking of a $\cal PT$-symmetric $S$ matrix, now with unbalanced gain and loss and even in the absence of gain. We show that the additional symmetry relation provides a clear evidence of an exceptional point, even when all other signatures of the $\cal PT$ symmetry breaking are completely erased. We also discuss the existence of a final exceptional point in this correspondence, which is attributed to asymmetric reflections from the two sides of the heterostructure.
\end{abstract}


\maketitle

Parity-Time ($\cal{PT}$) symmetric optical systems have attracted growing interest in the past few years. These systems are non-Hermitian due to the presence of gain and loss, which are delicately balanced such as the refractive index satisfies $n(x)=n^*(-x)$ with respect to a symmetry plane at $x=0$. The plethora of findings in such systems are tied to the spontaneous symmetry breaking at an exceptional point (EP) \cite{EP1,EP2,EPMVB,EP3,EP4,EP5,EP6,EP7,EP8,EP9,EP_CMT,PTlaser_nonlinear}. This spontaneous symmetry breaking was first suggested in non-Hermitian quantum mechanism \cite{Bender1,Bender2,Bender3} and later realized in the evolution of waves in the paraxial regime \cite{El-Ganainy_OL06,Moiseyev,Kottos,Musslimani_prl08,Makris_prl08,Feng}, which takes the system from a regime of real energy eigenvalues to complex conjugate pairs of eigenvalues. It has been shown that qualitatively similar behaviors exist even when such systems do not have exact $\cal PT$ symmetry, which leads to, for example, enhanced transmission with increased loss \cite{EP7} and reduced lasing emission with increased gain \cite{EP8,EP_CMT,PTlaser_nonlinear}.

Recently it was found that the scattering eigenstates of a $\cal PT$-symmetric system also display a spontaneous symmetry breaking \cite{CPALaser}, independent of its shape and dimension: the eigenvalues of the scattering ($S$) matrix can remain on the unit circle in the complex plane, conserving optical flux despite the non-Hermiticity; the symmetry breaking results in pairs of scattering eigenvalues with inverse moduli \cite{CPALaser,conservation}. However, unlike in previous studies, this symmetry breaking has been thought to vanish when gain and loss become unbalanced, with the exception of a very special case, i.e., the so-called ``mirror theorem" \cite{Robin}: the exceptional points of a $\cal PT$-symmetric heterostructure are unchanged when a pair of identical mirrors are attached symmetrically to its two sides. Although the mirrors can have gain or loss in them and hence break the overall $\cal PT$ symmetry of the system, the central region between the mirrors is still required to be $\cal PT$-symmetric.

In this report we reveal a symmetry property of the $S$ matrix eigenstates in a one-dimensional (1D) heterostructure, which is given by the amplitude ratio $\nu$ of the incident waves in the scattering eigenstates. We show that this symmetry relation is a result of the optical reciprocity \cite{Haus,Collin,Landau} and does not depend on the Hermiticy or the $\cal PT$ symmetry of the system. Using this relation, we demonstrate a quasi-transition of $\nu$ that is reminiscent of the aforementioned scattering $\cal PT$ symmetry breaking, now with unbalanced gain and loss and even in the absence of gain. We show that this additional symmetry relation provides a clear evidence of an exceptional point, even when all other signatures of the $\cal PT$ symmetry breaking are completely erased. Finally, we show the existence of a ``final exceptional point" in a multi-layer heterostructure, which is attributed to asymmetric reflections from the two sides of the heterostructure.

\begin{figure}[t]
\begin{center}
\includegraphics[width=\linewidth]{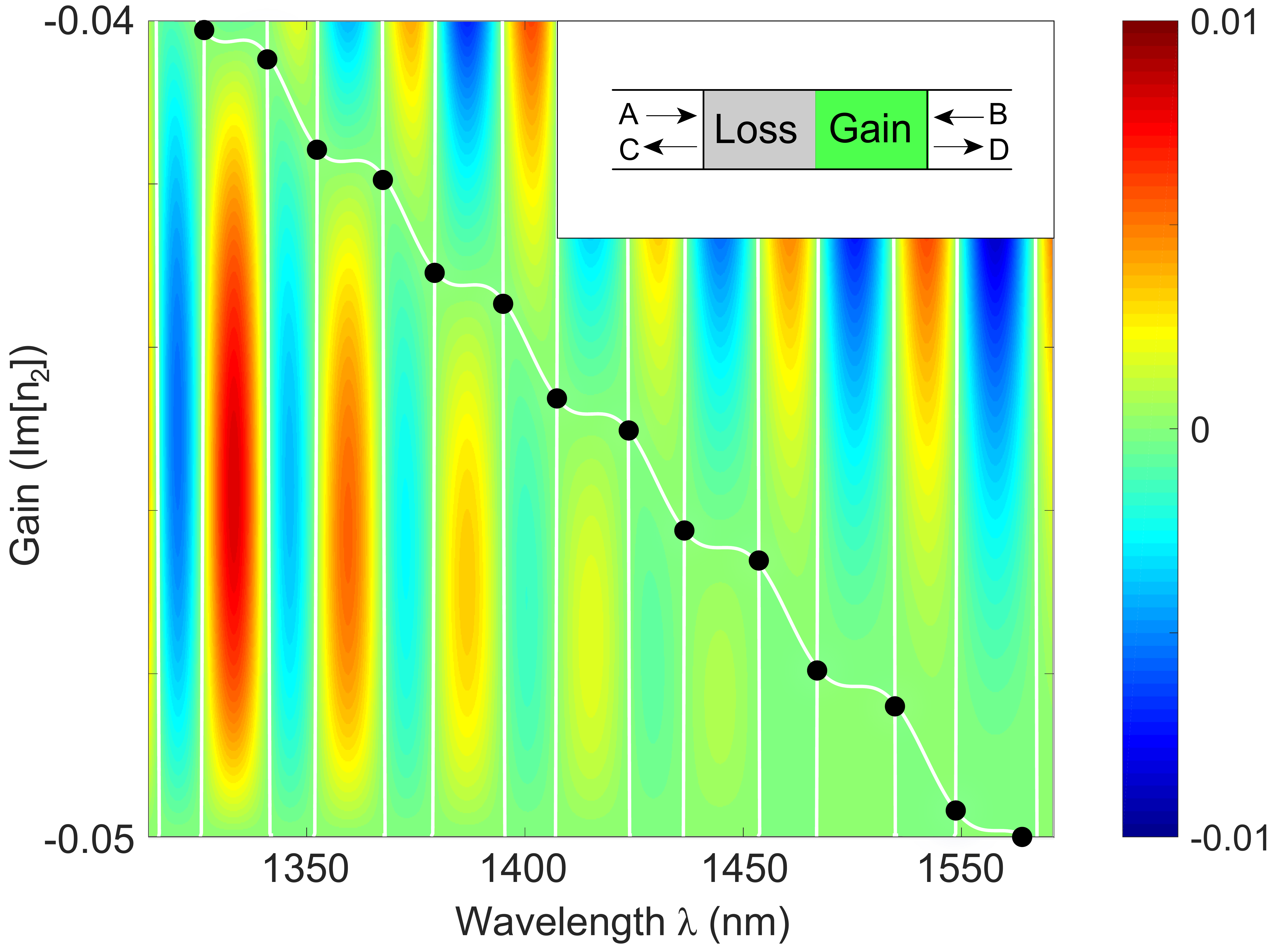}
\caption{(Color online) Exceptional points in a two-layer heterostructure (black dots) with weak unbalanced gain and loss. Loss is fixed at $\im{n_1}=0.05$ in the left half and the gain is reduced from $\im{n_2}=-0.05$ to $-0.04$ in the gain half. $\re{n}=3$ in the system and its length is $L=23\,\mu{m}$. False color plot of the product $\re{{\cal G}-i}\im{{\cal G}-i}$ is also shown. Nearly vertical and wavy diagonal lines show the zeros of $\re{{\cal G}-i}$ and $\im{{\cal G}-i}$, respectively. Their intersections show the locations of the exceptional points. ${\cal G}=-i$ does not hold in this region. Inset: Schematic of scattering from a two-layer heterostructure.}
\label{fig:imbalance_EP}
\end{center}
\end{figure}

Before we introduce this optical reciprocity induced symmetry property, it is worth reviewing the spontaneous symmetry breaking of the $S$ matrix in a 1D $\cal PT$-symmetric heterostructure. The $S$ matrix connects the incident waves to the scattered waves (see the inset in Fig.~\ref{fig:imbalance_EP}), e.g.,
\be
\begin{pmatrix}
C \\
D\end{pmatrix}
=
\begin{pmatrix}
r_L & t \\
t & r_R
\end{pmatrix}
\begin{pmatrix}
A \\
B\end{pmatrix}
\equiv
S
\begin{pmatrix}
A \\
B\end{pmatrix},\label{eq:S}
\ee
where $t\equiv t_L=t_R$ and $r_{L,R}$ are the transmission and reflection coefficients from the left and right sides.
Using the parametrization introduced in Ref.~\cite{conservation}, i.e.,
\be
S
=\frac{1}{a}
\begin{pmatrix}
ib & 1 \\
1 & ic
\end{pmatrix},
\ee
the eigenvalues of the $S$ matrix are given by
\be
\sigma_\pm = \frac{i}{2a}\left[(c+b) \pm \sqrt{(c-b)^2-4}\right].\label{eq:vas}
\ee
When the system is $\cal PT$-symmetric, $b,c$ are two real parameters and $a$ is complex parameter. They satisfy $|a|^2-1=bc$, which is another way of writing the generalized conservation law $|T-1|=\sqrt{R_LR_R}$ \cite{conservation}, where $T\equiv|t|^2,R_{L,R}=|r_{L,R}|^2$ are the transmittance and reflectances. When $|c-b|<2$, one finds $|\sigma_\pm|=1$ and the $S$ matrix is in the $\cal PT$-symmetric phase; when $|c-b|>2$, one finds that $\sigma_\pm$ have the same phase angle but their moduli are no longer 1.

Another manifestation of the spontaneous $\cal PT$ symmetry breaking, which is more relevant for the additional symmetry of the $S$ matrix we will introduce shortly, is exhibited in the amplitude ratios of the incident waves from the left and right sides (i.e., $\nu=A/B$) in the two scattering eigenstates. They are given by
\be
\nu_{\pm} = \frac{i}{2}\left[(c-b) \pm \sqrt{(c-b)^2-4}\right],\label{eq:ves}
\ee
which display the same qualitative change as $\sigma_\pm$ when the value of $|c-b|$ crosses 2. The latter is an exceptional point, at which $\sigma_\pm$ coalesce and so do $\nu_\pm$. We note that this condition for an exceptional point, as well as both Eq.~(\ref{eq:vas}) and (\ref{eq:ves}), holds even when the system is \textit{not $\cal PT$-symmetric}, in which case $a,b,c$ are three complex parameters in general.

Now let us return to the $\cal PT$-symmetric case. It can be easily checked that $|\sigma_+\sigma_-|=1$ and
\be
|\nu_+\nu_-|=1. \label{eq:con_ves}
\ee
These two relations, however, have very different origins. On the one hand, $|\sigma_+\sigma_-|=1$ holds only when $|a^2|-1=bc$, i.e., it is due to $\cal PT$ symmetry and breaks down when the gain and loss become unbalanced. $|\nu_+\nu_-|=1$, on the other hand, only requires that the $S$ matrix is symmetric, i.e., it is a result of the optical reciprocity \cite{Haus,Collin,Landau}. Since the optical reciprocity holds in general and does not rely on $\cal PT$ symmetry, $|\nu_+\nu_-|=1$ holds also with unbalanced gain and loss and even in the absence of gain.

$|\nu_+\nu_-|=1$ is the symmetry relation on which we rely to probe the reminiscence of the spontaneous $\cal PT$ symmetry breaking of the $S$ matrix when the system no longer has $\cal PT$ symmetry. We start by considering the simplest case, a heterostructure with two layers of equal width, the refractive index in which is $n_1$ and $n_2$, respectively.
The analytical expression of $S$ is given by
\be
\bm{S}=
\frac{1}{\cal D}
\begin{pmatrix}
{\cal G} + i{\cal F} & 1 \\
1 & -{\cal G} + i{\cal F}
\end{pmatrix} \label{eq:S_2layer},
\ee
where ${\cal D}\equiv c_1c_2 - gs_1s_2 - i(h_1s_1c_2 + h_2s_2c_1)$, ${\cal G}\equiv qs_1s_2$, ${\cal F}\equiv u_1s_1c_2 + u_2s_2c_1$, and $g=(n_1/n_2+n_2/n_1)/2$, $q=(n_1/n_2-n_2/n_1)/2$, $h_j = (n_j+1/n_j)/2$, $u_j = (n_j-1/n_j)/2$, $s_j=\sin(n_j\omega L_j/2c)$, $c_j=\cos(n_j\omega L_j/2c)\,(j=1,2)$. $\omega$ is the frequency of the incident light, and we note that $s_j,c_j$ are complex if $n_j$ is complex, i.e., when there is gain or loss. The eigenvalues of this $S$ matrix are given by
\be
\sigma_\pm = \frac{i{\cal F} \pm \sqrt{1+{\cal G}^2}}{\cal D}, \label{eq:sigma_2layer}
\ee
which indicates that if there is an exceptional point, then it occurs at
\be
{\cal G}=\pm i, \label{eq:EP}
\ee
where the radicand in Eq.~(\ref{eq:sigma_2layer}) vanishes.

In the $\cal PT$-symmetric case we have $n_1=n_2^*\equiv n+i\tau$, and it is straightforward to show that the $S$ matrix given by Eq.~(\ref{eq:S_2layer}) satisfies ${\cal PT}S{\cal PT} = S^{-1}$ \cite{CPALaser}, or simply $PS^*P=S^{-1}$, using $s_1=s_2^*$, $c_1=c_2^*$, $h_1=h_2^*$, $u_1=u_2^*$, $\re{q}=0$, and $\im{g}=0$. The superscript ``$^*$" denotes the complex conjugate as usual, and $P=\left(\begin{smallmatrix}0 & 1\\ 1 & 0 \end{smallmatrix}\right)$ is the matrix representation of the parity operator $\cal P$; it exchanges the incoming/outgoing waves on the left side of the heterostructure to those on the right side.
We also note that ${\cal G}=in\tau|s_1|^2/(n^2+\tau^2)$ is purely imaginary in this case, and hence the above condition (\ref{eq:EP}) for an exceptional point is reachable even if only one system parameter is varied, in contrast to the general requirement of sweeping at least a two-dimensional parameter space in non-$\cal PT$ systems \cite{EP2}. This property guarantees the two distinct phases of the $S$ matrix.

In the absence of the $\cal PT$ symmetry, the $S$ matrix still has exceptional points for complex values $n_1$ and $n_2$, since it is a non-Hermitian matrix \cite{EP2}. For example, in Fig.~\ref{fig:imbalance_EP} we show a two-layer heterostructure with fixed loss ($\im{n_1}=0.05$) in one half and weakly unbalanced gain ($\im{n_2}$) in the other. We found that its exceptional points are given by ${\cal G}=i$ in this regime, which exist at discrete pairs of $\{\im{n_2},\lambda\}$. $\lambda=2\pi c/\omega$ is the wavelength in vacuum, and its value at the exceptional points reduces with $\im{n_2}$. Since $\cal G$ now is complex in general and has an arbitrary phase angle, it no longer leads to two distinct phases of the $S$ matrix.

\begin{figure}[t]
\begin{center}
\includegraphics[width=\linewidth]{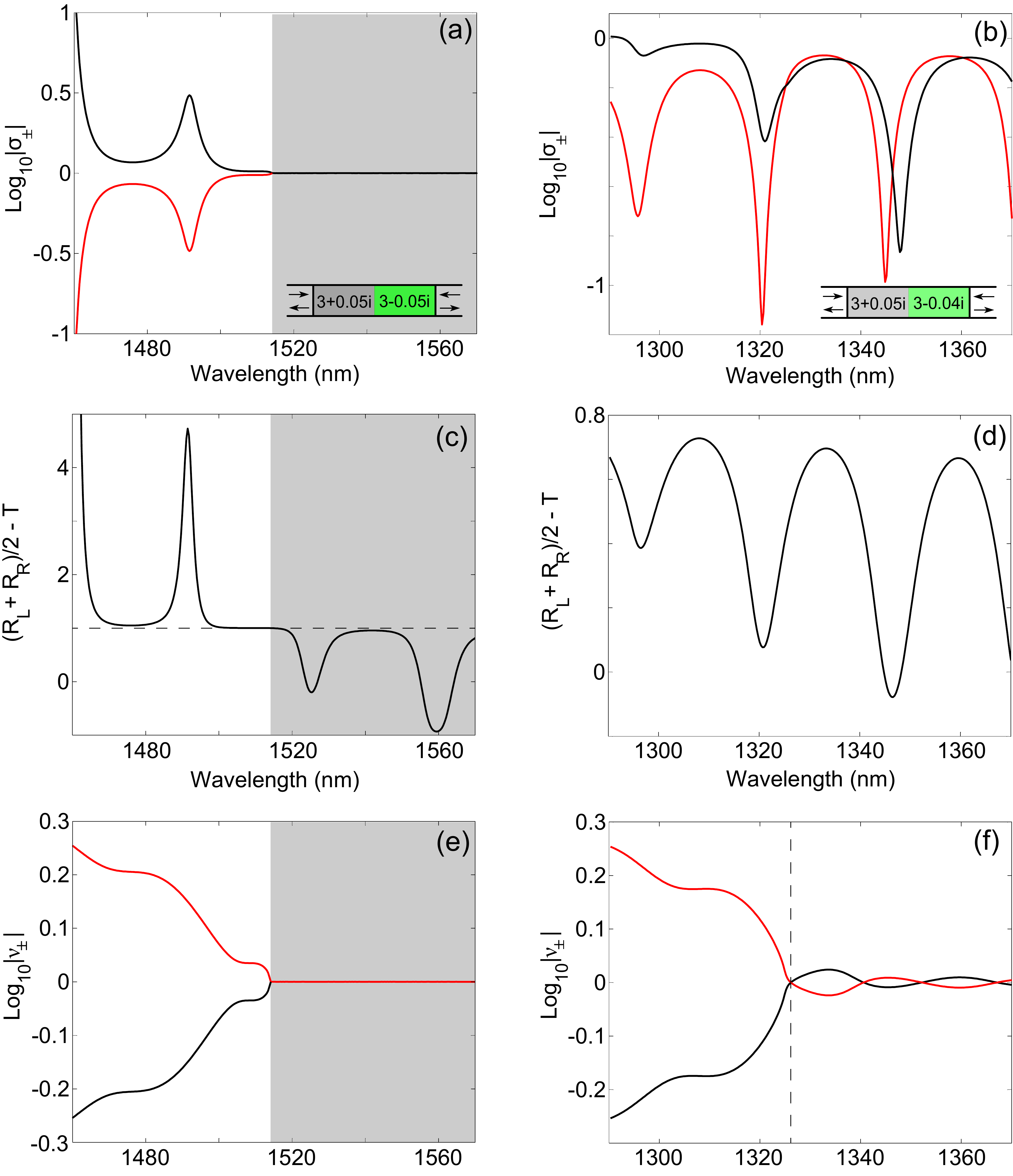}
\caption{(Color online) Contrast of the scattering behaviors when gain and loss are balanced (a,c,e; $\im{n_2}=-\im{n_1}=-0.05$) or weakly unbalanced (b,d,f;$\im{n_1}=0.04$ and $\im{n_2}=-0.05$). $\re{n}=3$ and $L=23\,\mu{m}$ as in Fig.~\ref{fig:imbalance_EP}. Shadowed areas in (a,c,e) indicate the broken symmetry phase. Dashed line in (d) marks the wavelength of the closet exceptional point at \{$\im{n_2}=-0.0401$, $\lambda$=1326 nm\} shown in Fig.~\ref{fig:imbalance_EP}.}
\label{fig:imbalance_ves}
\end{center}
\end{figure}

To be more specific, we note that $|\sigma_\pm|$ display a bifurcation at an exceptional point when the system is $\cal PT$-symmetric (see Fig.~\ref{fig:imbalance_ves}(a)), satisfying the relation $|\sigma_+\sigma_-|=1$ mentioned previously. This behavior no longer exists when there is a weak imbalance between gain and loss (see Fig.~\ref{fig:imbalance_ves}(b)), letting alone the case in which there is only loss in the heterostructure. Another indication of the spontaneous $\cal PT$ symmetry breaking is the transition of the difference $(R_L+R_R)/2-T$ from sub-unitary to super-unitary at an exceptional point (see Fig.~\ref{fig:imbalance_ves}(c)), which was derived using $|c-b|=2$ at an exceptional point in Eq.~(\ref{eq:vas}) and the $\cal PT$ symmetry relation $|a|^2-1=bc$ mentioned previously \cite{conservation}. This signature is also erased completely even when the gain and loss are weakly unbalanced (see Fig.~\ref{fig:imbalance_ves}(d)).

Now using the symmetry relation (\ref{eq:con_ves}) of the scattering eigenstates, the spontaneous $\cal PT$ symmetry breaking of the $S$ matrix can also be visualized as a bifurcation of $|\nu_\pm|$ (see Fig.~\ref{fig:imbalance_ves}(e)): they are equal in the $\cal PT$-symmetric phase and reciprocal of each other in the broken-$\cal PT$ phase. This behavior survives qualitatively when there is a weak imbalance between gain and loss, as we show in Fig.~\ref{fig:imbalance_ves}(f). We note that the quasi-transition point shown in Fig.~\ref{fig:imbalance_ves}(f) moves to a shorter wavelength with $\im{n_2}=-0.04$ when compared with the $\cal PT$-symmetric case (where $\im{n_2}=-0.05$). This is due to the blueshift of the exceptional point with reduced gain mentioned above (see Fig.~\ref{fig:imbalance_EP}). We can also check explicitly that the symmetry relation (\ref{eq:con_ves}) holds here: it is easy to convince oneself that the eigenstates of the $S$ matrix given by Eq.~(\ref{eq:S_2layer}) are the same as those of $\left(\begin{smallmatrix}{\cal G} & 1\\ 1 & -{\cal G} \end{smallmatrix}\right)$, and we find
\be
\nu_\pm = {\cal G} \pm \sqrt{{\cal G}^2+1};
\ee
their product is indeed $-1$.

\begin{figure}[t]
\begin{center}
\includegraphics[width=\linewidth]{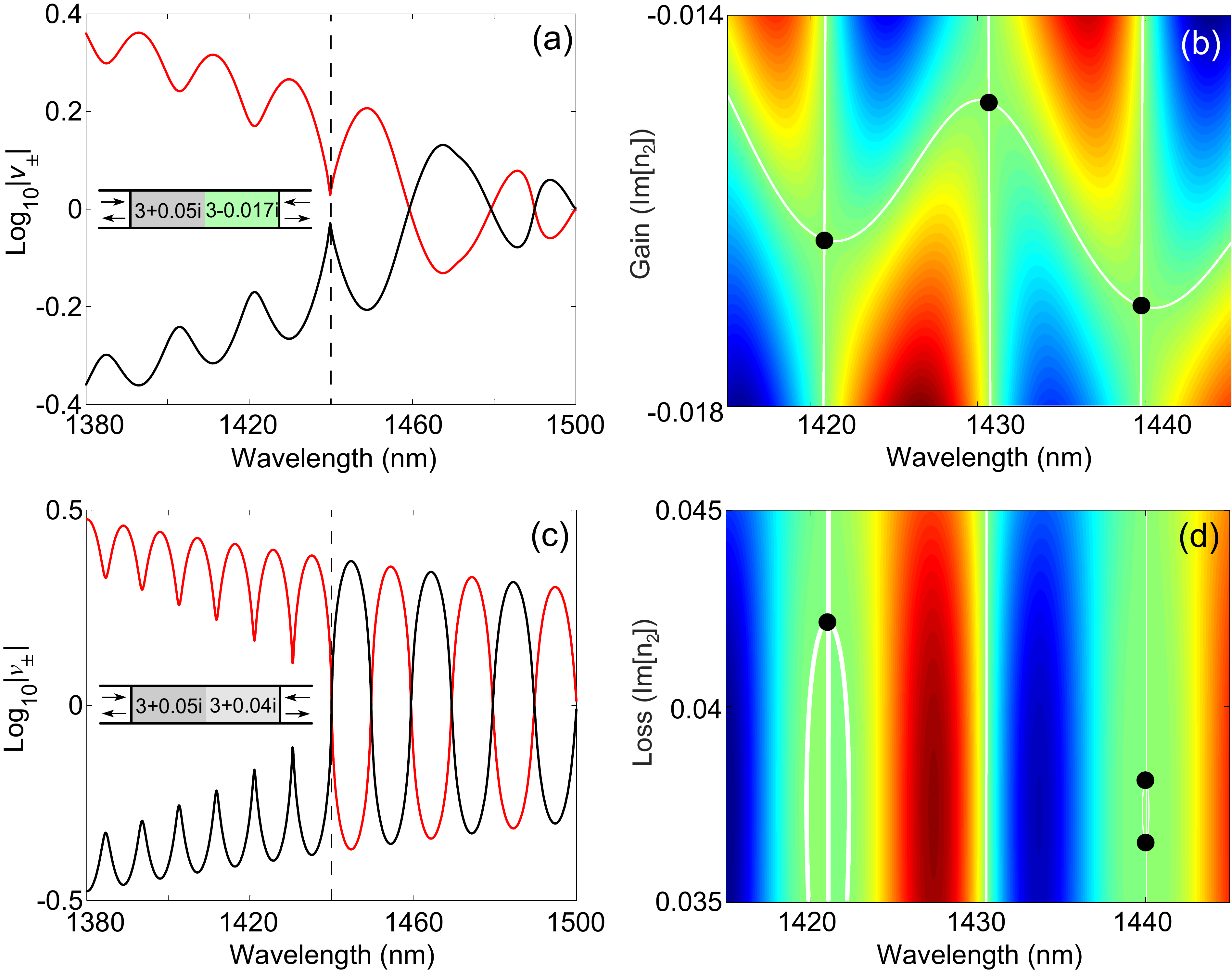}
\caption{(Color online) Amplitude ratios $\nu$ in the scattering eigenstates of the $S$ matrix with strongly unbalanced gain and loss (a) and loss only (c). $\im{n_2}=-0.17$ in (a) and 0.04 in (c), and $\im{n_1}=0.05$ is fixed.
Dashed lines in (a) and (c) mark the wavelength of the closet exceptional point at \{$\im{n_2}=-0.017$, $\lambda$=1440 nm\} and \{$\im{n_2}=0.038$, $\lambda$=1440 nm\}, respectively (see (b) and (d)).
System length is chosen to be $L=36\,\mu$m and the other parameters are the same as in Fig.~\ref{fig:imbalance_EP}. (b,d) Similar to Fig.~\ref{fig:imbalance_EP} for (a) and (c). In (d) ${\cal G}-i$ is replaced by ${\cal G}+i$.}
\label{fig:loss_ves}
\end{center}
\end{figure}

As the imbalance between gain and loss increases, so does the amplitude of the oscillations of $|\nu_\pm|$ shown Fig.~\ref{fig:imbalance_ves}(f). They weaken the distinctiveness of the quasi-transition but do not smear out the latter completely (see Fig.~\ref{fig:loss_ves}(a) at $\im{n_2}=-0.17$, for example). Interestingly, this observation holds even if the system only has loss, i.e., with both $\im{n_1},\im{n_2}>0$. In Fig.~\ref{fig:loss_ves}(c) we show the case when $\im{n_2}=0.04$, where $|\nu_\pm|$ approach each other and become interwoven beyond an exceptional point. We note that this exceptional point is now given by ${\cal G}=-i$, instead of ${\cal G}=i$ in the quasi-$\cal PT$ symmetric case shown in Figs.~\ref{fig:imbalance_EP} and \ref{fig:loss_ves}(b). There is one special point at which the quasi-transition of $|\nu_\pm|$ vanishes completely, that is when $\im{n_1}=\im{n_2}$. The system at this point is parity symmetric about the center of the heterostructure ($x=0$), and the two scattering eigenstates are even and odd functions of $x$, i.e., $|\nu_\pm|$ is always 1. We note that exceptional points in a loss-only system was previously studied in transmission \cite{EP7} and reflection \cite{Feng} experiments.

Next we discuss heterostructures with more than two layers. If the additional layers are identical and attached symmetrically to the two sides of the central region, we find that $\nu_\pm$ do not change their values and hence the quasi-transition of $|\nu_\pm|$ persists, no matter whether the additional layers have gain or loss.
This can be shown analytically by generalizing the ``mirror theorem" in $\cal PT$-symmetric heterostructures \cite{Robin}. For this purpose we utilize the transfer matrix $M$, which is defined by
\be
\begin{pmatrix}
C \\
A\end{pmatrix}
=
\frac{1}{t}
\begin{pmatrix}
{t^2-r_Lr_R} & {r_L} \\
-{r_R} & 1
\end{pmatrix}
\begin{pmatrix}
B \\
D\end{pmatrix}
\equiv
M
\begin{pmatrix}
B \\
D\end{pmatrix}\label{eq:M}
\ee
using the same notations as in Eq.~(\ref{eq:S}) for the central region.
Likewise, a transfer matrix $M_L$ and $M_R$ can be defined for the added left and right layers, and here they satisfy $PM_LP=M_R^{-1}$, where $P$ is the same matrix representing the parity operator introduced before. The total transfer matrix of the system with the mirrors is then given by $M'=M_L M M_R$. As Eq.~(\ref{eq:ves}) shows, $\nu_\pm$ of the central $\cal PT$-symmetric region only depend on $(c-b)$, or equivalently $\Delta \equiv (r_L-r_R)/t$, which is the sum of the two off-diagonal elements of $M$ in Eq.~(\ref{eq:M}). Therefore, to prove that $\nu_\pm$ do not change with the added mirrors, we only need to show that the sum of the two off-diagonal elements of $M'$, denoted by $\Delta' \equiv (r'_L-r'_R)/t$, equals $\Delta$. It is straightforward to show that
$\Delta' = \det(M_R)\Delta$. Since the determinant of a 1D transfer matrix is 1 in general \cite{Yeh}, this result concludes our proof.

\begin{figure}
\begin{center}
\includegraphics[width=\linewidth]{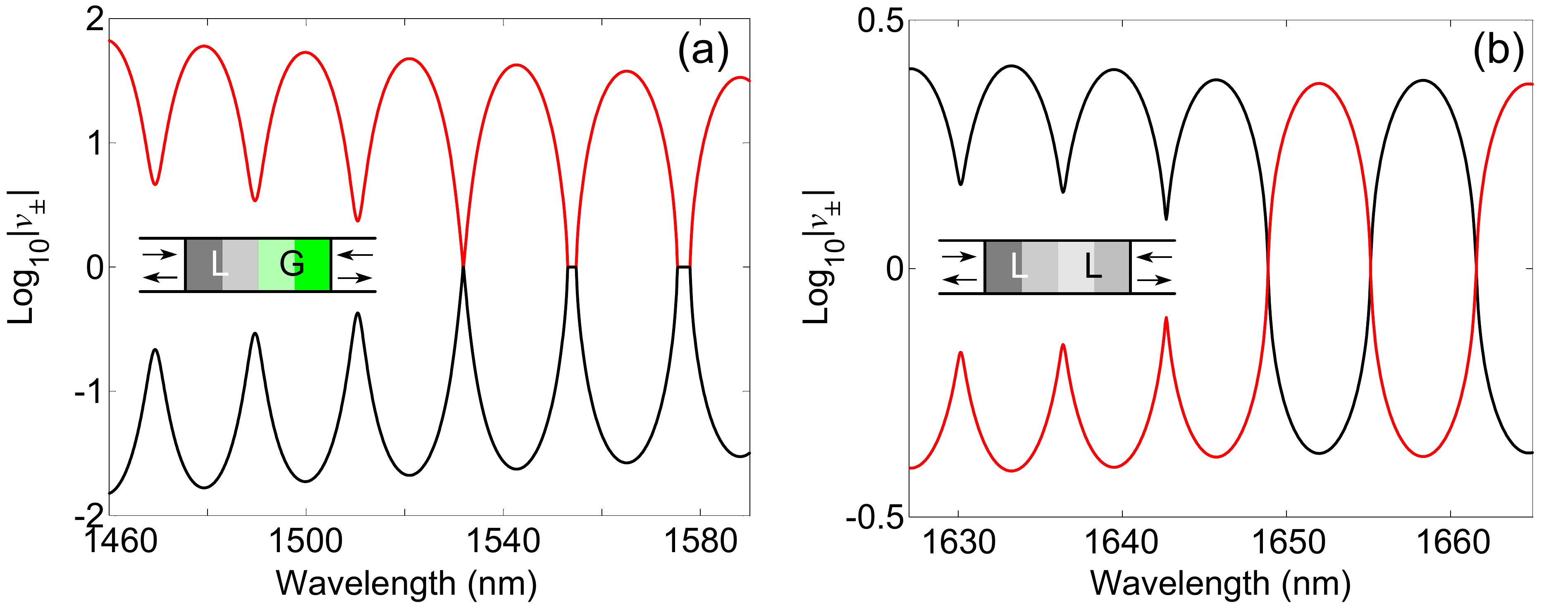}
\caption{(Color online) Amplitude ratios $\nu$ in the scattering eigenstates of the $S$ matrix in a 4-layer heterostructure. $\re{n}=3$ and $L=72\,\mu$m, and the four layers have equal length. (a) $\im{n}=0.05,0.01,-0.01,-0.05$ for a $\cal PT$-symmetric heterostructure. (b) $\im{n}=0.05,0.01,0.005,0.025$ for a loss-only waveguide.}
\label{fig:imbalance_complex}
\end{center}
\end{figure}

When the two layers added are different, the $S$ matrix of the $\cal PT$-symmetric system has multiple regions of symmetric and broken symmetry phases in general \cite{CPALaser}, each bounded by two exceptional points. The separations of these exceptional points in terms of wavelength are comparable to the oscillation periods of $|\nu_\pm|$ and can be fairly close. Hence these oscillations become more detrimental and obscure the bifurcations of $|\nu_\pm|$. However, in the strong gain/loss limit of a $\cal PT$-symmetric heterostructure, achieved with either a large $\tau$, a short wavelength, or a long system size, there seems to be a ``final" exceptional point, beyond which the system stays in the broken symmetry phase (see Fig.~\ref{fig:imbalance_complex}(a)). The existence of this final bifurcation point persists with unbalanced gain and loss and even in the absence of gain (see Fig.~\ref{fig:imbalance_complex}(b)), similar to the simplest two-layer waveguide discussed above.

This final exceptional point provides a good opportunity to gain a deeper understanding of the correspondence between the scattering behaviors in $\cal PT$-symmetric and non-$\cal PT$ heterostructure. As we have discussed, the exceptional points of the $S$ matrix is given by $c-b=\pm2$ in Eqs.~(\ref{eq:vas}) and (\ref{eq:ves}), or equivalently, $r_L-r_R=\pm2it$. In a $\cal PT$-symmetric heterostructure, $r_L$ and $r_R$ are in phase if $|t|<1$ and $\pi$ out-of-phase otherwise \cite{conservation}. When combined with a different form of the generalized conservation law, i.e., $|t|^2-1=-r_L^*r_R=-r_Lr_R^*$,
the above condition for the exceptional points becomes
\begin{align}
||r_L|-|r_R||&=2|t|, \, |r_L|+|r_R|=2 \hspace{17pt}(\text{if}\;|t|<1),\label{eq:d_amp} \\
||r_L|-|r_R||&=2, \hspace{11pt} |r_L|+|r_R|=2|t|\hspace{8pt} (\text{if}\; |t|>1)\label{eq:d_phase}.
\end{align}
For all the final exceptional point in $\cal PT$-symmetric heterostructures, including those in Fig.~\ref{fig:imbalance_ves}(e) and \ref{fig:imbalance_complex}(a), we always find the first scenario above (i.e., Eq.~(\ref{eq:d_amp})) to be true, which indicates a significant difference of $|r_L|,|r_R|$ when compared with $|t|$. In other words, it is this asymmetric reflection that leads to the final broken phase of the $S$ matrix in terms of the wavelength. Such asymmetric reflection does occur when the system is not $\cal PT$-symmetric, for example, when one half of the system has loss and the other half has unbalanced gain, or when the two halves have different average losses. This is especially the case in the short wavelength or large system limit, where the reflection from one side does not ``see" the other side of the system and $|t|\rightarrow 0$.

In conclusion, we have shown that the optical reciprocity leads to the symmetry relation $|\nu_+\nu_-|=1$, which holds in all 1D heterostructure. It manifests as a bifurcation of $|\nu_\pm|$ in $\cal PT$-symmetric systems when the spontaneous symmetry breaking of the $S$ matrix takes place, and this bifurcation persists qualitatively for the final exceptional point with unbalanced gain and loss and even in the absence of gain. Since tuning into the scattering eigenstates requires comparing the amplitudes \textit{and} phases of the scattered waves to those of the incident waves, measuring $\nu_\pm$ directly in the scattering eigenstates is rather inconvenient. One alternative is to measure $\nu_\pm$ indirectly using Eq.~(\ref{eq:ves}), with $b,c$ replaced by $-ir_L/t,-ir_R/t$.


We thank Douglas Stone for helpful discussions. L.G. acknowledges support by NSF under grant No. DMR-1506987. L.F. acknowledges support by NSF under grant DMR-1506884.
\bibliographystyle{longbibliography}

\end{document}